\documentclass[fleqn,twoside]{article}
\usepackage{espcrc2}

\usepackage{graphicx}

\newcommand{\AmS}{{\protect\the\textfont2
  A\kern-.1667em\lower.5ex\hbox{M}\kern-.125emS}}

\title{Monocular UHECR Spectra as Measured by HiRes}

\author{D.~R.~Bergman
  \address[Rutgers]{Rutgers, 
    The State Univeristy of New Jersey  \\
    Department of Physics and Astronomy \\
    Piscataway, New Jersey, USA 08854}
  presented on behalf of the High Resolution 
    Fly's Eye Collaboration}

\begin{document}

\begin{abstract}
   We have measured the spectrum of UHE cosmic rays in monocular mode
   using separately both detectors the High Resolution Fly's Eye
   experiment.  We describe the two detectors and the basic methods of
   analysis, and we present our measured spectra.  We compare these
   spectra with that produced by an astrophysical source model with
   galactic and uniformly distributed extra-galactic sources.  We also
   compare our spectra to the spectra produced by the AGASA experiment.
\vspace{1pc}
\end{abstract}

\maketitle

\section{Detectors}

The HiRes experiment consists of two air-fluorescence detectors
separated by 12.6~km and located at the U.S. Army Dugway Proving
Ground in Utah.  One detector, HiRes-I, consists of 22 mirrors
covering 3--17$^\circ$ in elevation and nearly 360$^\circ$ in
azimuth\cite{kn:HiRes-INIM}.  It uses a sample-and-hold data
acquisition system to save the photomultiplier tubes' pulse height and
time information.  HiRes-I has been running since June of 1997.  The
second detector, HiRes-II, located 12.6 km away, consists of 42
mirrors, covers 3--31$^\circ$ in elevation, and uses a flash ADC
(FADC) system to save pulse height and time information from its
phototubes \cite{kn:HiRes-IINIM}.  The sampling period of the FADC
electronics is 100 ns.  HiRes-II has been operating since October of
1999.  The mirror units at each site are identical, with a 5 m$^2$
mirror collecting light onto an array of 256 phototubes.  Each
phototube covers about one square degree of the sky.

\section{Monocular Analysis}

The two detectors are designed to observe cosmic ray showers
steroscopically.  However, HiRes-I has been running for much longer
than HiRes-II and at energies below about 10$^{18}$ eV, events can
only be observed by one detector.  Thus we are presenting here spectra
from monocular analyses.

Determination of the shower geometry in monocular mode is done by
fitting the trigger times, $t_{i}$, of hits to the following function
of the angles, $\chi_{i}$, of hits above the horizon,within the event
plane:
\begin{equation}
t_{i}=t_{0}+\frac{R_{p}}{c}\tan{\left(\frac{\pi-\psi-\chi_{i}}{2}\right)}
\label{timefit}
\end{equation}
Here $R_{p}$ is the impact parameter, $\psi$ the in-plane angle
between the shower and the horizon, and $t_{0}$ the time of closest
approach. 

With limited elevation coverage, HiRes-I monocular events are too
short in angular spread for reliable determination of $\psi$ and
$R_{p}$ by timing alone. For this analysis, the expected form of
shower development itself was used to constrain the time fit to yield
realistic geometries.  The shower profile was assumed to be described
by the Gaisser-Hillas parameterization
\begin{equation}
N(x) = N_{m}
\left(\frac{x-x_{0}}{x_{m}-x_{0}}\right)^
{(x_{m}-x_{0})/\lambda}
e^{(x_{m}-x)/\lambda}
\label{gh}
\end{equation}
where $N(x)$, $N_{m}$ are the number of particles at depth $x$ and at
shower maximum depth $x_{m}$, respectively.  The first-interaction
depth and shower elongation constant are denoted by $x_{0}$, and
$\lambda$~\cite{gh}.This technique is called the profile-constrained
fit (PCF). Equation~\ref{gh} is in agreement with previous HiRes
measurements~\cite{hr_prof}, and with CORSIKA/QGSJET
simulations~\cite{csong,corsika,qgsjet}. Based on these, we fixed
$x_{0}$ and $\lambda$ at 40 and 70~g/cm$^{2}$, respectively. We
allowed $x_{m}$ to vary in 35~g/cm$^{2}$ steps between 680 and
900~g/cm$^{2}$, matching the expected range for proton to iron
primaries in this energy range.  This procedure breaks down for events
with energies less than $3\times{10}^{18}$~eV, providing the lower
limit to the HiRes-I spectrum.

The greater elevation coverage of HiRes-II allows for the
reconstruction of the shower geometry from timing alone.
Equation~\ref{timefit} is linear in $R_p$ and $t_0$, we find the best
value for these variables analytically for each $\psi$; the best
$\psi$ is then found by $\chi^2$ minimization.  With the geometry of
the shower known, we fit the observed light signal to the
Gaisser-Hillas parameterization of Eqn.~\ref{gh}. We collected
photo-electrons from all tubes into a sequence of time bins.  This
exploited the FADC data acquisition system and lessened our
sensitivity to PMT acceptance.

Two calibration issues effect both sites: the absolute calibration of
the phototube gains and the determination of the transparency of the
atmosphere.  The phototube gains are determined through a nightly
illumination of the phototube clusters with a YAG laser and a monthly
illumination by a Xenon flash bulb.  The atmospheric transparency is
determined by probing the atmosphere hourly with a set of laser shots
from each of the sites which is observed and reconstructed by the
other site.

\section{Data-MC Comparisons}

\begin{figure}[htb]
  \includegraphics[width=7.5cm]{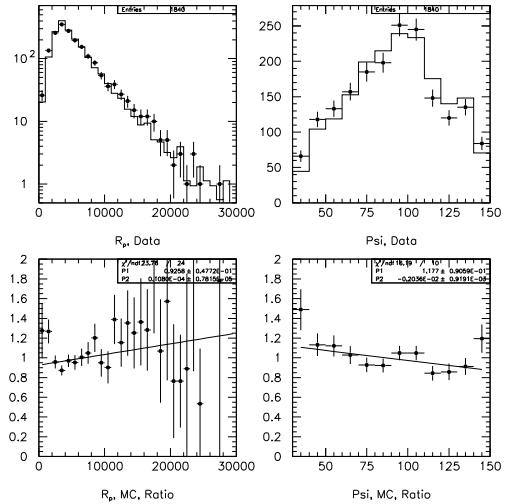}
  \caption{Data/MC comparisons of $r_p$ and $\psi$.}
  \label{fig:MCgeo}
\end{figure}

\begin{figure}[htb]
  \includegraphics[width=7.5cm]{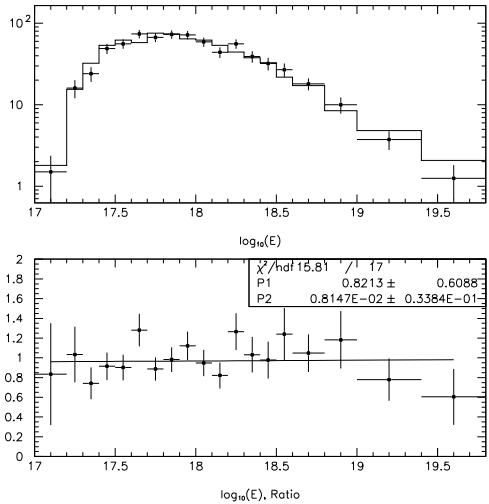}
  \caption{Data/MC comparison of reconstructed energies.}
  \label{fig:MCE}
\end{figure}

The aperture as a function of energy is a necessary component of the
spectrum calculation.  It is determined by Monte Carlo calculation.
To verify our calculated aperture and analysis procedures, we
undertook an extensive program of Monte Carlo simulation development
and comparison of simulated data with real data.  Two figures
exemplify this; both are from the HiRes-II analysis.  In
figure~\ref{fig:MCgeo} we compare data and MC distributions of two
geometric variables: $r_p$ and $\psi$.  In figure~\ref{fig:MCE} we
compare the distributions of reconstructed energies.  Note that any
large discrepancy between the atmospheric parameters used in MC and
the conditions under which the data were actually taken would heavily
influence the distribution of impact parameters.

\section{The HiRes Spectra}

\begin{figure}[htb]
  \includegraphics[width=7.5cm]{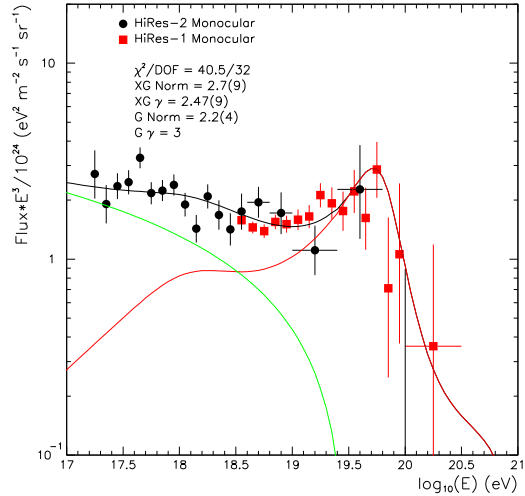}
  \caption{HiRes-I and HiRes-II spectra fit to galactic and uniform
    extragalactic source model.}
  \label{fig:hrspectra}
\end{figure}

The two HiRes monocular spectra are shown in
figure~\ref{fig:hrspectra} \cite{HR1,HR2} along with what one would
expect from a model combining galactic sources with uniformly
distributed extra-galactic sources\cite{Berezinsky}.  The GZK cutoff is
evident, and there is no need from this data to invoke extraordinary
means in order to evade it.

\begin{figure}[htb]
  \includegraphics[width=7.5cm]{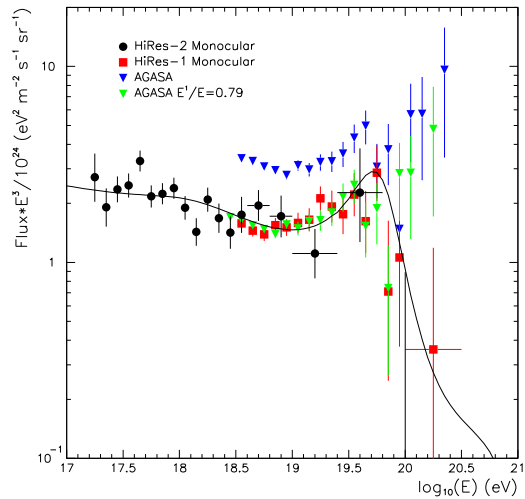}
  \caption{HiRes-I and HiRes-II spectra compared with AGASA.}
  \label{fig:AGASA}
\end{figure}

In figure~\ref{fig:AGASA} we compare our spectra with that of
AGASA\cite{AGASA}.  The shape of the two spectra agree quite well for
energies below 100 EeV, and reducing the AGASA energies by 20\%, which
is within AGASA's stated systematic uncertainty, makes this comparison
explicit.  The famed discrepancy between HiRes and AGASA has thus been
reduced to a few points above 100 EeV, where AGASA has five events
(with they're energy scale reduced by the 20\%) and HiRes has only
one.

\end{document}